\documentclass[a4paper,fleqn]{cas-dc}

\usepackage[numbers]{natbib}
\usepackage{tabularx}
\usepackage{hhline}
\usepackage{pifont}

\usepackage{caption}
\usepackage{subcaption}
\usepackage[normalem]{ulem}
\usepackage{xcolor}

\def\tsc#1{\csdef{#1}{\textsc{\lowercase{#1}}\xspace}}
\tsc{WGM}
\tsc{QE}

\begin{document}
\let\WriteBookmarks\relax
\def\floatpagepagefraction{1}
\def\textpagefraction{.001}

\shorttitle{FDG PET image synthesis from T1 MRI. Application to unsupervised deep brain anomaly detection}    

\shortauthors{Zotova, Pinon, Trombetta, Bouet, Jung, Lartizien}  

\title[mode = title]{GAN-based synthetic FDG PET images from T1 brain MRI can serve to improve performance of deep unsupervised anomaly detection models}  



%

\author[1]{Daria Zotova}





\credit{Methodology, Software, Validation, Formal analysis, Writing - Original Draft}


\affiliation[1]{organization={INSA Lyon, Université Claude Bernard Lyon 1, CNRS, Inserm, CREATIS UMR 5220, U1294},
            city={Lyon},
            postcode={F-69621}, 
            country={France}}

\author[1]{Nicolas Pinon}[orcid=0000-0001-8817-9911]

\credit{Methodology, Software, Writing - Original Draft, Writing - Review \& Editing, Visualization}

\author[1]{Robin Trombetta}[orcid=0009-0001-7945-6953]

\credit{Writing - Original Draft, Writing - Review \& Editing, Visualization}

\author[2]{Romain Bouet}[orcid=0000-0002-3690-2976]

\credit{Data Curation, Writing - Review \& Editing}

\author[2]{Julien Jung}

\credit{Validation, Resources, Data Curation, Writing - Review \& Editing}

\author[1]{Carole Lartizien}[orcid=0000-0001-7594-4231]
\cormark[1]
\ead{carole.lartizien@creatis.insa-lyon.fr}

\credit{Conceptualization, Methodology, Validation, Formal analysis, Data Curation, Writing - Original Draft, Writing - Review \& Editing, Supervision, Project administration, Funding acquisition}





\affiliation[2]{organization={Lyon Neuroscience Research Center, INSERM U1028, CNRS UMR5292, Univ Lyon 1},
            city={Bron},
            postcode={69500}, 
            country={France}}

\cortext[1]{Corresponding author}




\begin{abstract} 
\textbf{Background and Objective.} Research in the cross-modal medical image translation domain has been very productive over the past few years in tackling the scarce availability of large curated multi-modality datasets with the promising performance of GAN-based architectures. However, only a few of these studies assessed task-based related performance of these synthetic data, especially for the training of deep models. \textbf{Method.} We design and compare different GAN-based frameworks for generating synthetic brain [18F]fluorodeoxyglucose (FDG) PET images from T1 weighted MRI data. We first perform standard qualitative and quantitative visual quality evaluation. Then, we explore further impact of using these fake PET data in the training of a deep unsupervised anomaly detection (UAD) model designed to detect subtle epilepsy lesions in T1 MRI and FDG PET images. We introduce novel diagnostic task-oriented quality metrics of the synthetic FDG PET data tailored to our unsupervised detection task, then use these fake data to train a use case UAD model combining a deep representation learning based on siamese autoencoders with a OC-SVM density support estimation model. This model is trained on normal subjects only and allows the detection of any variation from the pattern of the normal population. We compare the detection performance of models trained on 35 paired real MR T1 of normal subjects paired either on 35 true PET images or on 35 synthetic PET images generated from the best performing generative models. Performance analysis is conducted on 17 exams of epilepsy patients undergoing surgery. \textbf{Results.} The best performing GAN-based models allow generating realistic fake PET images of control subject with SSIM and PSNR values around 0.9 and 23.8, respectively and \textit{in distribution} (ID) with regard to the true control dataset. The best UAD model trained on these synthetic normative PET data allows reaching 74\% sensitivity. \textbf{Conclusion.} Our results confirm that GAN-based models are the best suited for MR T1 to FDG PET translation, outperforming transformer or diffusion models. We also demonstrate the diagnostic value of these synthetic data for the training of UAD models and evaluation on clinical exams of epilepsy patients. Our code and the normative image dataset are available.
\end{abstract}




\maketitle

\section{Introduction}

One major limitation to the performance of supervised deep learning models for medical image analysis is the scarce availability of large annotated training databases. 
The tasks of accurate image labeling and sampling of the pathological patterns variability, indeed, constitute the main bottlenecks to the collection of large datasets since it is time-consuming and relies on clinical expertise which is severely time-constraint. 
For this reason, unsupervised, weakly supervised, or few shot learning paradigms have gained significant interest over the past few years, due to the constraint release on the annotation process \cite{Tajbakhsh_MEDIA20}. This includes anomaly detection models, which were shown to perform well, especially for detection and segmentation tasks in neuroimaging \cite{baur_MEDIA2020}. This subgroup of methods which consists of learning normal representations or patterns extracted from normal (i.e. non-pathological) populations only, allows relaxing the pressure on the annotation of the pathological cases. 
However, gathering large datasets of normative populations is another challenge, since the vast majority of available clinical images are patient data with pathological patterns. 
A general dilemma arises when a certain modality is available for a patient dataset but is missed for the healthy population needed for effective training of anomaly detection models. 
The targeted acquisition of such a large normative database is also not to be considered due to stringent regulation and cost issues, especially for more invasive techniques, such as nuclear imaging modalities, like positron emission tomography.\\
The objective of this study is to build on the recent cross-modality synthesis architectures to efficiently generate realistic brain [18F]fluorodeoxyglucose (FDG) PET images derived from T1 MRI images of a normal subjects population. Research in this specific cross-modal translation domain has been very productive over the past few years for the synthesis of pathological data for dementia (e.g. Alzheimer disease) \cite{armanious2019unsupervised,sikka2018mri,wei2019predicting,wolterink2017generative,yaakub2019pseudo,Pan_TMI20}, mainly evaluating the visual quality of the synthetic PET data with standard metrics such as Peak Signal-to-Noise Ration (PSNR) or Structural similarity index metric (SSIM). Among all proposed architectures, including the most recent transformer or denoising diffusion models, adversarial models such as GAN or Cycle-GAN remain best performers for this specific tasks \cite{Dayarathna_MEDIA24}. To the best of our knowledge, no study has examined the quality of normative synthetic PET data, and only a few studies assessed diagnostic task-based related performance of synthetic data, especially in the context of unsupervised anomaly detection \cite{yaakub2019pseudo,Burgos_JIMAGING21}.\\

\begin{figure}[h]
    \centering
    \includegraphics[width=8cm]{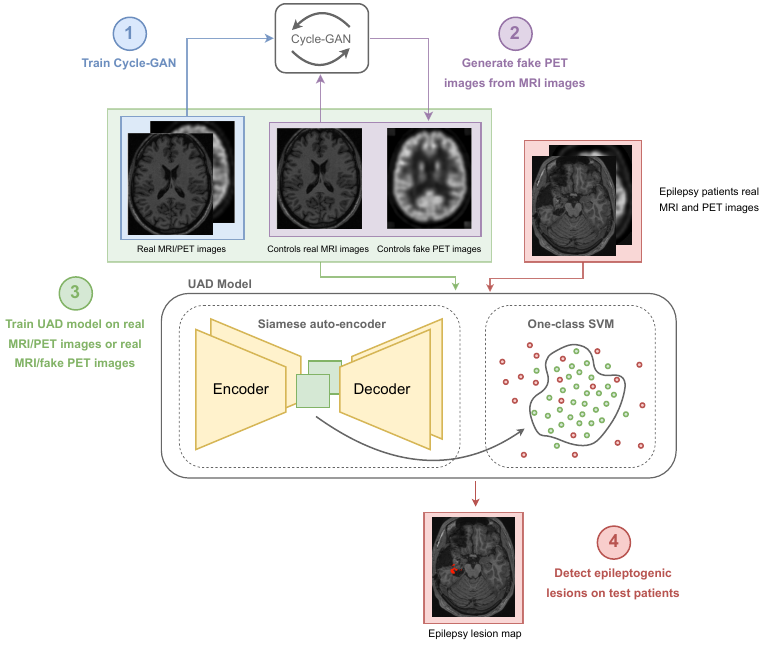}
    \caption{\label{fig:pipeline} Graphical abstract depicting the main steps of the conducted analysis.}
\end{figure} 

The main contributions of this paper, depicted on Figure \ref{fig:pipeline}, are:
\begin{itemize}
    \item A model derived from the Cycle-GAN architecture for the synthesis of realistic FDG PET exams of normal subjects from T1 MRI.
    \item A fair comparison of different variants of GANs architectures based on the same training and validation dataset and the same evaluation metric.
    \item A global evaluation of the quality of these synthetic data including standard quantitative metrics of visual image quality as well as novel task-oriented OOD metrics.
    \item A use case application demonstrating their ability to mimic real FDG PET normative data for the training of a deep unsupervised anomaly detection model applied to detect subtle brain anomalies in clinical bimodal T1 MRI and FDG PET exams.
\end{itemize}

This paper builds on our preliminary work \cite{Zotova_Sashimi21}. We first significantly extend and update the state-of-the-art analysis. Then we introduce diagnostic task-oriented quality metrics tailored to UAD models and  of the synthetic FDG PET imaging data and strengthen the use case application focusing on the deep unsupervised anomaly detection of epileptogenic zone (EZ) in multimodal FDG PET and T1 MRI exams of a series of epilepsy patients 
who underwent surgery and had good outcomes based on Engel's classification. \cite{Wieser_Epilepsia01} (Engel I or II).

\section{Related work}
\label{sec:related_work}
This section reviews the state-of-the-art bibliography in the domain of cross-modal medical image synthesis whose aim is to generate one imaging from one or more other modalities data with the main objective of tackling the issue of missing data. We specifically highlight translation studies from T1 MRI to FDG PET imaging with a specific focus on the visual and task-oriented image quality metrics implemented in the state-of-the-art studies of this domain. 

In \cite{Dayarathna_MEDIA24}, the authors comprehensively review deep learning-based medical imaging synthesis studies from 2018 to 2023 including a specific section on MRI to PET cross-modal translation.
Research in this specific cross-modal image translation has been very productive over the past few years. 
One of the first studies presented by \cite{sikka2018mri} aimed at synthesizing FDG PET from MRI exams of normal control (NC) and Alzheimer (AD) patients for data augmentation purposes. Performance of a 3D U-Net architecture was evaluated on a subset of exams extracted from the ADNI dataset based on standard image quality metrics, such as structural similarity index metric (SSIM). The authors also included some preliminary diagnostic task-based evaluations measuring the added value of the generated synthetic PET  images in the training of a classification network to discriminate patient from normal subjects exams. A few years later, the same group experimented a GAN-based architecture based on the same experimental paradigm in \cite{Sikka_arivx2021}. Studies based on the ADNI dataset were also conducted at the same period in \cite{Pan_TMI20} and \cite{Pan_TPAMI22} based on the Cycle-GAN architecture for PET image synthesis and NC vs AD diagnostic downstream task. In \cite{Pan_MICCAI21}, the same group also performed a first attempt to jointly train the generative network synthesizing the missing modality and the classification network.  In \cite{Zhang_CMPB22}, the authors explored performance of a 3D GAN based architecture trained on ADNI normal, MCI and AD patients, both including quantitative visual metrics as well as classification performance. More recently, \cite{xie_arxiv24} proposed a joint diffusion and attention model (JDAM) allowing to generate FDG PET from T1 MRI. This model was trained on 2D transverse slices of MRI and PET exams of ADNI patients coregistered to the MNI atlas. Visual quality metrics indicated similar PSNR and slightly higher SSIM values that those reported on synthetic PET exams generated by a standard cycleGAN. Radiologists' qualitative evaluation of synthetic PET images generated by CycleGAN and JDAM indicated no noticeable differences. As also reported by \cite{Dayarathna_MEDIA24}, all these studies demonstrate good performance of GAN-based architectures for this specific cross-modal translation both in terms of visual metrics and added value of the generated synthetic PET data to augment training datasets of classification networks, most predominantly targeting the discrimination of AD patients.

While most of these studies focused on synthesizing pathological PET exams of AD patients for data augmentation purpose and patient-level diagnostic downstream tasks, only a few studies addressed the need to generate normative PET image datasets to improve the training of unsupervised anomaly detection models, which is the main topic of interest of our study. In \cite{yaakub2019pseudo}, a 3D-patch-based approach for FDG PET synthesis from normal T1 MRI was proposed based on a GAN model with a residual U-Net as a generator. This model trained on exams of healthy subjects only was shown to outperform two other models based on simple 3D U-Net and high-resolution dilated CNN in terms of mean absolute error (MAE) and peak signal-to-noise ratio (PSNR). 
The authors also used this generative model to synthesize control-like pseudo-FDG PET images from MR T1 exams of epilepsy patients, which were then subtracted from the patient's true FDG PET images to localize hypometabolic regions. This method was shown to outperform standard statistical parametric mapping (SPM) analysis.

A first conclusion from this synthetic overview of the literature is that GAN-based architectures remain among the best state-of-the-art models, outperforming more recent architectures based on transformers or diffusion models, for the specific tasks of T1 MRI to FDG PET cross-modal translation. 

Another conclusion is that most of the recent papers in the domain include, in addition to standard quality metrics such as PSNR or SSIM, performance on downstream tasks, comparing the performance of the clinical task at hand with and without synthetic data. 
These studies mostly focus on evaluating the added value of synthetic PET for the training of classification models discriminating AD from NC or MCI patients, based on the ADNI dataset. We found no other study than \cite{yaakub2019pseudo} that addressed segmentation or detection tasks with synthetic FDG PET data derived from T1 MRI. Another way to evaluate that synthetic data follow the same distribution as the reference true data is to evaluate if the synthetic data are out-of-the-distribution (ODD) samples. A few OOD detection methods have been proposed over the past few years, mostly for deep classification networks. In \cite{GONZALEZ_MEDIA22}, a lightweight method was recently proposed to detect when deep segmentation models silently fail on OOD data, mainly due to domain shift problems. This method exploits the Mahalanobis distance in the feature space of the deep segmentation model to derive uncertainty maps which are shown to correctly signal when a model prediction should not be trusted. It demonstrated good performance for different segmentation tasks including the Covid-19 lung CT lesion segmentation challenge gathering multi-center scans as well as MRI segmentation tasks of the hippocampus and the prostate. Such OOD analysis has not been performed both in the context of UAD and based on synthetic PET data. Our objective in this study is to leverage these recent task-oriented performance metrics to the analysis of synthetic brain PET data generated from T1 MRI based on efficient Cycle-GAN architectures.

\section{Method}
\subsection{GAN architectures for the synthesis of realistic FDG PET data from T1 MRI}
\label{sec:GAN_models}

As stated in section \ref{sec:related_work}, generative adversarial networks (GANs) \cite{goodfellow2014generative, wolterink_radiographics21} have demonstrated impressive results for cross-modal T1 MRI to FDG PET medical image translation \cite{Dayarathna_MEDIA24}.
The basic structure of a GAN consists of the generator that is trained to generate new synthetic samples and the discriminator that tries to distinguish examples being real or fake. These two models are trained simultaneously and compete against each other.
In this study, we build on a comparative analysis of different variants of GAN architectures to design the optimal configuration for missing FDG PET data generation from T1 MRI data. More precisely, we compare standard GAN and Cycle-GAN architectures with two types of input data, consisting either of adjacent full transverse slices or 3D patches sampled throughout the 3D image volumes, and with adapted loss terms. 

\subsubsection{GANs architectures\\}

\begin{figure}[h]
    \centering
    \includegraphics[width=8cm]{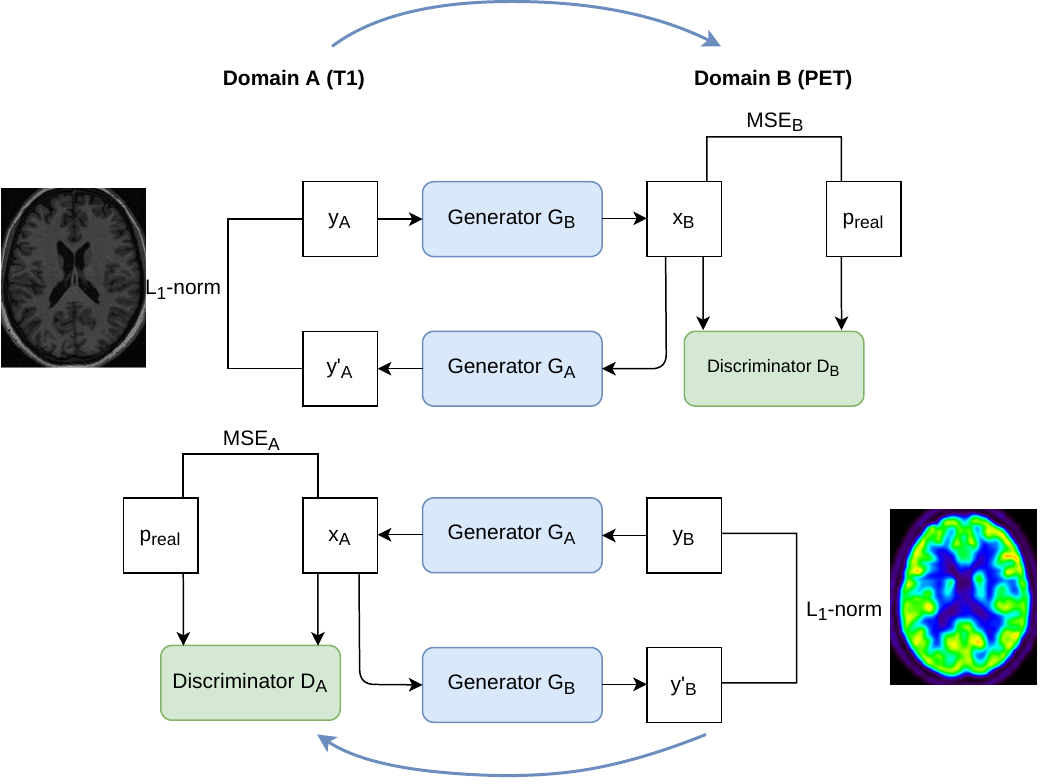}
    \caption{\label{fig:CycleGAN}Cycle-GAN architecture based on two baseline GANs translating images from domain A to domain B (upper GAN) and vice versa (lower GAN). 
    }
\end{figure}

\emph{Simple-GAN}. We first propose to use a standard GAN architecture with one generator $G_B$ and one discriminator $D_B$ as depicted in the upper part of Figure \ref{fig:CycleGAN}.
Generator $G_B$ attempts to improve the quality of the translated output $x_b$ of domain $B$ from the original input $y_A$ from domain $A$, thus deceiving the discriminator $D_B$. 
The training procedure is formulated as a min-max optimization problem of an objective function that the discriminator is trying to maximize and the generator is trying to minimize. 
In this study, we implement the least squares GAN (LSGAN) model \cite{mao2017LSGAN} that aims to minimize the following discriminator $L_{LSGAN}(D_B,A,B)$ and generator $L_{LSGAN}(G_B,A,B)$ losses : 

\begin{equation} 
\begin{split} 
L_{LSGAN}(D_B,A,B) & = \mathbb{E}_{p(x_b)}[D_B(x_b)^2] +  \mathbb{E}_{p{(y_b)}}[(D_B(y_b)-1)^2] \\
L_{LSGAN}(G_B,A,B) & = \mathbb{E}_{p(x_b)}[(D_B(x_b)-1)^2]
\label{Eq_LSGAN}
\end{split}
\end{equation}
where $y_a$ and $y_b$ are true images of domain A and B, respectively, and $x_b=G_B(y_a)$ is the fake image of domain B generated from $y_a$. \\

\emph{Cycle-GAN}. Cycle-GAN consists of two generator networks $G_A$ and $G_B$ and two discriminator networks $D_A$ and $D_B$. The baseline Cycle-GAN model is shown in Figure \ref{fig:CycleGAN}. The generators translate images from domain A to domain B and vice versa. Each of the generator networks is trained adversarially using a corresponding discriminator $D_A$ and $D_B$. In addition to the adversarial loss term of the simple GAN network in eq. (\ref{Eq_LSGAN}), the key element in training the Cycle-GAN network is a cycle-consistency loss function $L_{cyc}$:

\begin{equation}
\label{Eq_CycleLoss}
L_{cyc}(G_A,G_B)=\mathbb{E}_{p(y_a)}[\parallel y_a'-y_a\parallel_1]+\mathbb{E}_{p(y_b)}[\parallel y_b'-y_b\parallel_1]
\end{equation}

where $y_a'$ is the fake image of domain A generated by generator $G_A$ from the fake $x_b$, that is $y_a'= G_A(x_b)$ with $x_b= G_B(y_a)$.

\subsubsection{Adapted loss term for supervised image translation}

CycleGAN-models have been proposed and shown efficient when trained on unpaired modalities, e.g. T1 MRI and FDG PET acquired from different patients. In the context of supervised image translation, where the model can be trained on paired images in both domains at the pixel level, which means corresponding images of the same patient, we propose to add a mean squared error (MSE) loss term $L_{mse}$ between the fake image $x_b$ generated from a true image $y_a$ of domain A and its paired true image $y_b$ in domain B:

\begin{equation}
\label{Eq_MSEloss}
L_{mse}(G_B)=\mathbb{E}_{p(x_b) }[\parallel x_b-y_b\parallel_2^2]
\end{equation}

This MSE loss term between real and synthetic images of both domains A and B can be added to the simple GAN loss formulation in eq (\ref{Eq_LSGAN}) as well as to the Cycle-GAN one in eq (\ref{Eq_CycleLoss}).

\subsubsection{Image-based versus patch-based models and post-processing}
Two-dimensional slice-based models taking one single slice as input allow capturing the global spatial context but inherently fail to leverage context from adjacent slices, unlike 3D models which can lead to improved performance but come with a heavy computational and data cost. 
As a compromise, we consider two configurations depending on the size and shape of the input data:
\begin{itemize}
    \item \textbf{\textit{2.5D}} models which receive three adjacent transverse slices as input (each slice corresponding to one channel) and output the corresponding three adjacent FDG PET slices of same dimension as the input ones.
    \item \textbf{\textit{3D-patch}} models where we feed 3D mini patches extracted from the original 3D images into the network and output the corresponding 3D FDG PET patch of same dimension as the input one.
\end{itemize}

In the 2.5D configuration, inference of a synthetic 3D FDG PET image is performed by first splitting the 3D T1w MRI input volume in non-overlapping triplets of adjacent slices, which are inputted to the trained 2.5 model, and then stacking the generated transverse slices to reconstruct the synthetic 3D FDG PET image. In the 3D configuration, inference of a synthetic 3D FDG PET image is performed by first splitting the 3D T1w MRI input volume in overlapping 3D patches with a stride equal to half the patch size in 3 dimensions, which are inputted to the trained 3D-patch model. The 3D PET volume is reconstructed by cropping the generated 3D patches of half the patch size in each dimension and stacking these cropped patches. This crop allows discarding the edge pixels with lower prediction confidence as shown in \cite{huang2018tiling}.
 For both 2.5D and 3D-patch configurations, we finally apply Gaussian smoothing as a post-processing to tackle "border" effects that may occur when stacking either slices or mini-volumes. As a lightweight normalization, we also perform standard histogram matching by adjusting the intensity distribution of any fake PET to that of a randomly chosen PET image of the database that served to train the GAN synthesis model and considered as the reference image.

\subsection{Visual quality metrics.}
\label{sec:visual-metric}
Quantitative score of translated images compared with the true image as reference were estimated based on the following standard visual quality metrics : Mean Squared Error (MSE), Peak Signal-to-Noise Ratio (PSNR), Structural Similarity Index Metric (SSIM) and  Learned Perceptual Image Patch Similarity (LPIPS). A definition of these metrics is provided in Appendix \ref{sec:AppendixA}. \\

\subsection{Task-oriented out-of-distribution (OOD) detection quality metrics. }
\label{sec:OOD-metric}

As stated in section \ref{sec:related_work}, we are interested in going beyond visual quality assessment and estimating the quality of the synthetic data to perform a task of interest. In this study, the synthetic PET data mimicking the distribution of FDG in the normal population serve to train a brain T1 MRI and FDG PET UAD model for the task of epileptogenic zone screening. The description of the UAD model considered in this study is described in section \ref{sec:brain_ano_cad}.
The main principle of such autoencoder-based UAD models is to model the normal population by learning to compress and recover data from healthy subjects. Once trained, anomalies present in pathological data can be detected as outliers from the modeled normative distribution, either by computing reconstruction error between the true and reconstructed images \textit{in the image space} or by training generative or density support estimation (e.g. OC-SVM) models \textit{in the latent representation space}.

To evaluate if the distribution of the generated synthetic PET data matches that of true PET data, we derive OOD detection metrics fitted to the specific unsupervised anomaly detection task at hand.

The first metric is the global MSE (as defined in eq. \ref{MSE_eq} of Appendix A) averaged over all pixel-based error between any input image $u$ (either fake or true) to the autoencoder AE and its reconstruction $AE(u)$. In case the autoencoder takes multiple images (e.g. PET and T1 MRI) as input, the global MSE is the average metric computed for each input channel. This metric quantifies if the fake FDG PET data mimics in-distribution (ID) true FGD PET data \textit{in the image space}, thus validating the use of fake PET images to train UAD models based on the reconstruction error.

The second OOD metric, inspired by the idea developed in \cite{GONZALEZ_MEDIA22}, is derived from the computation of the Mahalanobis distance $D_{m}^{i}$ between the latent representation vector ${z_i}$ of any input element $i$ and the distribution of this latent variable in the normal training population data, as

\begin{equation}
\label{mahal_dist_eq}
{D_{m}^{i} = ({z_i}-\mu)^{T}\Sigma^{-1}({z_i}-\mu)}
\end{equation}

where $\mu$ and $\Sigma$ are the empirical mean and covariance, respectively, computed over the latent representation vectors $z_i$ of the N elements of the training data set as:

\begin{equation}
\label{mu_cov_eq}
\mu = \frac{1}{N}\sum_{i=1}^{N}z_i \qquad \Sigma = \frac{1}{N-1}\sum_{i=1}^{N}(z_i - \mu)(z_i-\mu)^{T}
\end{equation}

Following \cite{GONZALEZ_MEDIA22}, we  compute $D_{m}^{i}$ for each voxel of the 3D image, then average it over all voxels to obtain a global metric referred to as $D_m$, whose value is normalized between the minimum and doubled maximum values for ID train data. This global $D_m$ metric quantifies if the fake FDG PET data mimic \textit{in-distribution} (ID) true FGD PET data \textit{in the latent space}, thus validating the use of fake PET to train UAD models performing the detection task in the latent representation space. Also note, that when multiple images are inputted as channels to the autoencoder, the latent representation tensor $z$ will encode information from the different types of images. 

Comparing the distributions of MSE and Mahalanobis distance $D_m$ of true $y$ and fake $x$ images will allow detection of any domain shift in the image and latent spaces, respectively.

\subsection{Application to the training of a deep epilepsy lesion detection model}
\label{sec:brain_ano_cad}
Unsupervised anomaly detection (UAD) was proposed as a natural alternative to supervised frameworks. This formalism requires only the manual identification of "normal" data to construct a tractable model of normality, while outliers are automatically detected at inference time as samples deviating from this normal model.

Artificial neural networks have been extensively used for UAD, mainly based on standard autoencoder (AE) architectures \cite{baur_MEDIA2020} or on more advanced architectures, e.g. combining a vector quantized autoencoder (VQ-VAE) with autoregressive transformers \cite{pinaya_MEDIA22}. 
These AE models are trained to perform a ``pretext'' task on normal images consisting in the reconstruction of these images. At inference, voxel-wise anomaly scores of any arbitrary image are then computed as reconstruction errors, i.e. the differences between the original image and the pseudo-normal reconstructed one.  Such errors are expected to be much larger for unseen voxels from patient images, provided the chosen architecture has initially well captured the normal subjects' main features.
The models have been successfully applied to the segmentation of visible brain anomalies in MRI medical image analysis \cite{baur_MEDIA2020}. 

We recently compared different auto-encoder architectures for the detection of subtle anomalies in the diffusion parametric maps of \textit{de novo} Parkinson's disease (PD) patients \cite{munoz_mlcn2021} and for the detection of white matter hyperintensities (WMH) in T1 and FLAIR brain MRIs \cite{pinon_MIDL2023}. These studies confirmed recent observations outlining the limitations of the reconstruction error scores for the detection of subtle abnormalities \cite{meissen2022_pitfalls}.
To overcome these reported limitations, we proposed to perform the detection step in the latent space of the autoencoder by coupling the learned representation to a non-parametric discriminative one-class support vector machine (OC-SVM) \cite{alaverdyan_MEDIA2020}. This model was recently shown to achieve promising results for the detection of WMH lesions in T1 and FLAIR MRI \cite{pinon_MIDL2023}, outperforming reconstruction error models, as well as for the classification of \textit{de novo} PD patients versus controls \cite{pinon_isbi23}. In \cite{alaverdyan_MEDIA2020}, it was originally developed to detect subtle epilepsy lesions from T1 and FLAIR MRI images and showed promising performance with 62\%  lesion detection sensitivity in MRI-negative patients.

In this study, we adapt it to the  challenging detection task of subtle epileptogenic zones (EZ) based on T1 MRI and FDG PET images. FDG PET imaging, which can highlight the focal reduction of glucose metabolism at the EZ localization, is a routine clinical exam for the presurgical evaluation of epilepsy patients. As for MRI, abnormalities of the FDG signal may be challenging and time-consuming to detect visually. 
As stated in the introduction, setting up a normative dataset of T1 MRI and FDG PET exams of healthy subjects is a difficult task due to stringent regulation for more invasive techniques, such as positron emission tomography. 

\begin{figure*}[h]
\centering
\includegraphics[width=12cm]{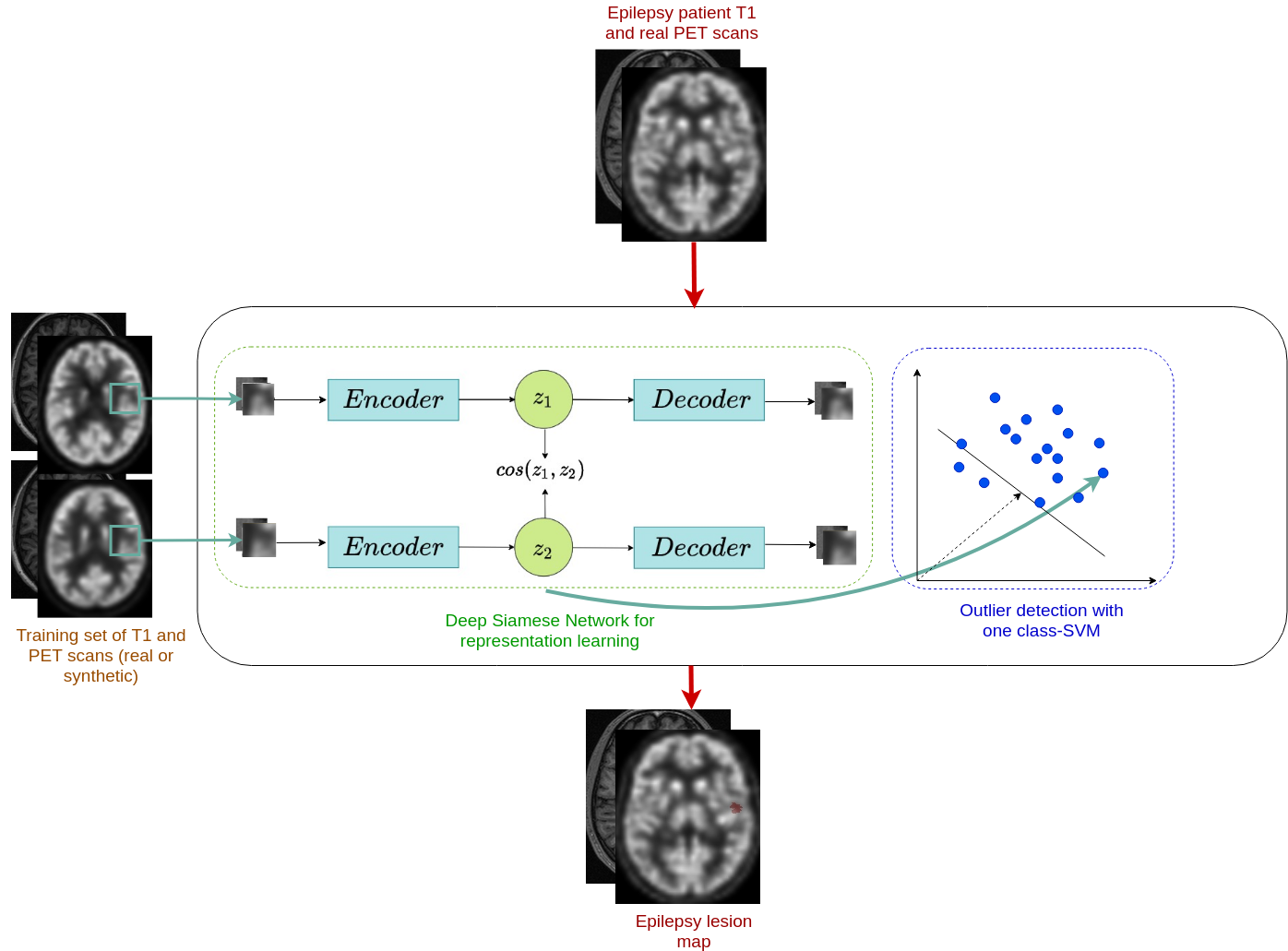}
\caption{\label{fig:brain_ano_CAD}Scheme of brain anomaly detection model proposed in \cite{alaverdyan_MEDIA2020}. The original model was trained to detect epilepsy lesion from T1 and FLAIR images. In this study, it is adapted to perform the detection task based on T1 and [18F]FDG PET images.}
\label{fig:ocsvm}
\end{figure*}

We propose to train the unsupervised anomaly detection (UAD) model of \cite{alaverdyan_MEDIA2020} with paired T1 MRI and true or fake FDG PET data to evaluate the task-oriented quality of the synthetic normative FDG PET data. This trained model then serves to detect EZ in a series of T1 MRI and FDG PET data of real patients.
As depicted in Figure \ref{fig:brain_ano_CAD}, this model encompasses a siamese network composed of stacked convolutional autoencoders which are designed to map patches extracted from a healthy control population to a latent representation space. Once training of this autoencoder is performed, we associate the latent representation yielded by this network to the central voxel of each inputted paired MR T1 and PET patches extracted from the 3D brain volume. These latent representations serve as feature vectors that are then fed into OC-SVM models \cite{scholkopf2001estimating}.
We train one OC-SVM model per voxel on the matrix composed of the latent representations of this voxel extracted from all subjects from the healthy control dataset. 
When tested on patients, this model allows outlining the locations (at the voxel level) of abnormalities with regards to the normal brain population, thus producing anomaly score maps, which are then thresholded to produce cluster maps. These clusters are ranked according to a criterion defined in \cite{alaverdyan_MEDIA2020}, which accounts for the cluster size and its average score. Appendix \ref{sec:AppendixC} provides details on the hyperparameters setting of the UAD model and post-processing of the anomaly score maps.

We thus propose to compare performance achieved by the UAD model of \cite{alaverdyan_MEDIA2020} in two training scenarios: in the first scenario, the model is trained on paired true FDG PET and T1 MRI images of normal subjects, while in the second, it is trained on the same true T1 MR images but paired with fake FDG PET data generated from the GAN-based architectures proposed in section \ref{sec:GAN_models}. In both scenarios, performance is evaluated on the same series of real T1 and FDG PET exams of epilepsy patients.

\section{Experiments and performance analysis}
\subsection{Data description and preprocessing}
\label{sec:data}

This study, including three clinical image databases, was approved by our institutional review board
with approval numbers 2012-A00516-37 and 2014-019 B and written consent was obtained for all participants.

\begin{itemize}
    \item The first database, $DB_{C1}$, consists of a series of 35 paired FDG PET and T1 weighted MRI scans co-registered to the MNI space, thus leading to 3D image volumes of size 157x189x136 with 1mm$^3$ isotropic voxel size. These data were acquired on 35 healthy volunteers with 19 women aged 23-65 years (mean age $38.5\pm 11.8$ years) on a 1.5T Sonata scanner and mCT PET scanner (Siemens Healthcare, Erlangen, Germany). They are fully described and made available to the community in \cite{Merida_EJNMI21}.
    \item The second database, $DB_{C2}$, consists of 40 T1-weighted MRI exams acquired on healthy control subjects aged 20-62 years with 20 women on the same 1.5T MR Sonata scanner, but different from those of $DB_{C1}$. 
    \item The third database, $DB_{ep}$, consists of 17 paired FDG PET and T1 weighted MRI scans of patients with 19 confirmed medically intractable and subtle epileptogenic zone (EZ) localization, as illustrated in Table \ref{tab:detection_res} (Patients $A$ to $Q$) and Figure \ref{fig:cluster_maps}. The EZ localization was indeed not visible on the T1 MRI (MRI negative) of all considered patients while it induced slight hypometabolism on about half of the considered FDG PET exams. One extra patient (Patient $R$) in Table \ref{tab:detection_res} with surgical resection of the right temporal lobe was also considered to evaluate the performance of the models in detecting non-epileptogenic and large abnormalities. The 18 patients with 11 women were aged 19-58 years (mean $35 \pm 10.8$ years) and all underwent epilepsy surgery. The localization of the epileptogenic zone was established by combining intracranial EEG data, surgical reports showing the location of cortectomy as well as surgical outcome based on Engel Classification \cite{Wieser_Epilepsia01}. We indeed kept patients who had a good outcome (Engel I-II). The located EZ was then manually outlined by expert clinicians. These data were acquired on the same 1.5T Sonata MR scanner and mCT PET scanner (Siemens Healthcare, Erlangen, Germany) and with the same parameter setting as those used for $DB_{C1}$ and $DB_{C2}$. 
\end{itemize}

The T1-weighted MRI volumes of all three databases were first processed with the unified segmentation algorithm implemented in SPM12 (\url{https://www.fil.ion.ucl.ac.uk/spm/doc/manual.pdf}) using the default parameter values. This algorithm performs tissue segmentation (white/grey matter, cerebrospinal fluid), correction for magnetic field inhomogeneities, and spatial registration to the standard brain template of the Montreal Neurological Institute (MNI) with a voxel size of 1×1×1 mm. PET images of $DB_{C2}$ and $DB_{ep}$ were then rigidly aligned to their corresponding individual T1w MR images in the native space. Then, they were co-registered to the MNI space with SPM12 applying the transformation parameters derived from the registration of the T1 images. A standard min-max normalization was finally applied to all 3D  FDG PET and T1w-MRI images.
A masking image in the MNI space derived from the Hammersmith maximum probability atlas described in \cite{Hammers_Atlas2003} was used at different steps of the pipeline, to focus or exclude specific brain regions.

For the 2.5D configuration described in section \ref{sec:GAN_models}, where the GAN architectures receive pseudo-3D images at input corresponding to three adjacent transverse slices, the original co-registered 3D images (of size 157x189x136 with 1 $mm^{3}$ isotropic voxel) were cropped to 128x128x136 3D image volumes centered on the brain to remove background voxels. 
For the 3D-patch configuration, the original 3D images were first resized to 3D images of size 160x192x160 with zero padding, and then sets of mini-volumes were extracted, each of size 32x32x32. 

\subsection{Overview of the conducted experiments}
\label{sec:exp_overview}
The main objective of this study is to demonstrate that we can generate normative FDG PET images from T1 MRI that not only look real but also can be used in combination with their paired true T1 MRI to train auto-encoder-based UAD models. To that purpose, we conduct three series of experiments that are summarized on Table \ref{tab:exp_summary}.

The first series of experiments consists of training the different deep GAN-based models introduced in section \ref{sec:GAN_models} and compare their performance based on visual quality metrics described in section \ref{sec:OOD-metric}. These experiments are conducted on the $DB_{C1}$ database (true T1 MRI and PET) in a cross-validation manner. The selected best 2.5D and 3D-patch GAN models from this first experiment are then fully retrained based on the $DB_{C1}$ exams considering 29 training, 3 validation and 3 test controls. Finally, the 40 T1 MRI of $DB_{C2}$ are inputted to these two models to generate two series of 40 synthetic FDG PET data, one from the best 2.5D model and one from the best 3D-patch model. The paired 40 true T1 MRI of $DB_{C2}$ and synthetic PET from the best 2.5D model form the $DB_{C2}^{2.5D}$ dataset while the same 40 true T1 MRI and synthetic PET from the best 3D-patch model form the $DB_{C2}^{3Dpatch}$ dataset.

The second series of experiments consists of computing task-oriented OOD metrics as described in section \ref{sec:OOD-metric}. To that purpose, we train the UAD model of Figure \ref{fig:brain_ano_CAD} on 35 out the 40 samples of the $DB_{C2}^{2.5D}$ and $DB_{C2}^{3Dpatch}$ datasets respectively, leading to $UAD_2$ and $UAD_3$ models. We then evaluate the $MSE$ and $D_m$ metrics for the 35 control subjects of $DB_{C1}$, for the remaining 5 test synthetic PET exams of $DB_{C2}^{2.5D}$ ($DB_{C2}^{3Dpatch}$) that were not used to train $UAD_2$ ($UAD_3$), as well as for the 18 patients of $DB_{ep}$.

The last series of experiments consists in evaluating the detection performance of the three UAD models, namely $UAD_2$ and $UAD_3$ as well as $UAD_1$ which is trained on the series of 35 paired true T1 MRI and FDG PET of $DB_{C1}$. This evaluation is conducted on the series of 18 patients of $DB_{ep}$, thus allowing a fair comparison between UAD models trained on true and fake paired T1 MR1 and FDG PET normative exams.

In the following, we provide details of these different experiments.

\begin{table*}
\caption{Synopsis of the three different experiments, with the dataset used at each step. $DB_{C2}^{Fake}$ is either $DB_{C2}^{2.5D}$ or $DB_{C2}^{3Dpatch}$ and is, in both cases, divided into 35 training samples and 5 test samples.}
\scriptsize
\def\arraystretch{1.5}
\begin{tabularx}{\textwidth}{|>{\raggedright\arraybackslash}m{5cm}|>{\raggedright\arraybackslash}m{2cm}|>{\centering\arraybackslash}m{0.7cm}|>{\centering\arraybackslash}m{0.7cm}|X|X|>{\centering\arraybackslash}m{0.7cm}|>{\arraybackslash}m{3.4cm}|}
    \hhline{--------}
    \multicolumn{8}{|c|}{\cellcolor[HTML]{DDDDDD}Experiment 1 : GANs comparison for PET synthesis} \\ \cline{1-8}
    \multicolumn{1}{|c|}{\multirow{2}{*}{Steps descriptions}} & \multicolumn{1}{c|}{\multirow{2}{*}{Model(s)}} & \multicolumn{1}{c|}{\multirow{2}{*}{$DB_{C1}$}} & \multicolumn{1}{c|}{\multirow{2}{*}{$DB_{C2}$}} & \multicolumn{2}{c|}{$DB_{C2}^{Fake}$} & \multicolumn{1}{c|}{\multirow{2}{*}{$DB_{ep}$}} & \multicolumn{1}{c|}{\multirow{2}{*}{Outputs/results}} \\ \cline{5-6}
     & & & & \multicolumn{1}{>{\centering\arraybackslash}p{0.7cm}|}{35} & \multicolumn{1}{>{\centering\arraybackslash}p{0.7cm}|}{5} & & \\ \cline{1-8}
     
    1) Train 2.5D and 3D-patch GAN models. 4-fold cross-validation and experimental details in Appendix \ref{sec:AppendixB}.  & Simple GAN or Cycle-GAN & \ding{51} & & \multicolumn{1}{>{\centering\arraybackslash}p{0.7cm}|}{} & \multicolumn{1}{>{\centering\arraybackslash}p{0.7cm}|}{} &  & Table \ref{tab:test_reslts} \\ \cline{1-8}
    
    2.a) Train the best 2.5D GAN model  & Cycle-GAN with MSE loss & \ding{51} & & & &  & 2.5D-GAN model \\ \cline{1-8}
    
    2.b) Train the best 3D-patch GAN model  & Cycle-GAN with MSE loss & \ding{51} & & & &  & 3Dpatch-GAN model \\ \cline{1-8}
    
    3.a) Generate $DB_{C2}^{Fake} = DB_{C2}^{2.5D}$ fake PET images & 2.5D-GAN & & \ding{51} & & &  & $DB_{C2}^{2.5D}$ \\ \cline{1-8}
    
    3.b) Generate $DB_{C2}^{Fake} = DB_{C2}^{3dpatch}$ fake PET images & 3Dpatch-GAN & & \ding{51} & & &  & $DB_{C2}^{3Dpatch}$ \\ \hhline{--------}

    \multicolumn{8}{|c|}{\cellcolor[HTML]{DDDDDD}Experiment 2 : Task-oriented synthesis models performance evaluation} \\ \cline{1-8}
    \multicolumn{1}{|c|}{\multirow{2}{*}{Steps descriptions}} & \multicolumn{1}{c|}{\multirow{2}{*}{Model(s)}} & \multicolumn{1}{c|}{\multirow{2}{*}{$DB_{C1}$}} & \multicolumn{1}{c|}{\multirow{2}{*}{$DB_{C2}$}} & \multicolumn{2}{c|}{$DB_{C2}^{Fake}$} & \multicolumn{1}{c|}{\multirow{2}{*}{$DB_{ep}$}} & \multicolumn{1}{c|}{\multirow{2}{*}{Outputs/results}} \\ \cline{5-6}
     & & & & \multicolumn{1}{>{\centering\arraybackslash}p{0.7cm}|}{35} & \multicolumn{1}{>{\centering\arraybackslash}p{0.7cm}|}{5} & & \\ \cline{1-8}
     
    1) Train UAD model on real MRI + real PET images  & UAD model of \cite{alaverdyan_MEDIA2020} & \ding{51} & & \multicolumn{1}{>{\centering\arraybackslash}p{0.7cm}|}{} & \multicolumn{1}{>{\centering\arraybackslash}p{0.7cm}|}{} &  & $UAD_1$ \\ \cline{1-8}
    
    2) Train UAD models on real MRI + fake PET images ($DB_{C2}^{2.5D}$ or $DB_{C2}^{3dpatch}$) & UAD model of \cite{alaverdyan_MEDIA2020} & & \ding{51} & \multicolumn{1}{>{\centering\arraybackslash}p{0.7cm}|}{\ding{51}} &  \multicolumn{1}{>{\centering\arraybackslash}p{0.7cm}|}{} &  & $UAD_2$ and $UAD_3$ \\ \cline{1-8}

    3) Compare true and fake PET images and evaluate the synthesis with task-oriented OOD metrics & $UAD_2$ and $UAD_3$ & \ding{51} & & \multicolumn{1}{>{\centering\arraybackslash}p{0.7cm}|}{}& \multicolumn{1}{>{\centering\arraybackslash}p{0.7cm}|}{\ding{51}} & \ding{51} & Figure \ref{fig:SyntheticPET} for examples \newline Figure \ref{fig:2.5D_3Dpatch_scatter_plot} for OOD measurements \\ \hhline{--------}

    \multicolumn{8}{|c|}{\cellcolor[HTML]{DDDDDD}Experiment 3 : Anomaly detection performance evaluation} \\ \cline{1-8}
    \multicolumn{1}{|c|}{\multirow{2}{*}{Steps descriptions}} & \multicolumn{1}{c|}{\multirow{2}{*}{Model(s)}} & \multicolumn{1}{c|}{\multirow{2}{*}{$DB_{C1}$}} & \multicolumn{1}{c|}{\multirow{2}{*}{$DB_{C2}$}} & \multicolumn{2}{c|}{$DB_{C2}^{Fake}$} & \multicolumn{1}{c|}{\multirow{2}{*}{$DB_{ep}$}} & \multicolumn{1}{c|}{\multirow{2}{*}{Outputs/results}} \\ \cline{5-6}
     & & & & \multicolumn{1}{>{\centering\arraybackslash}p{0.7cm}|}{35} & \multicolumn{1}{>{\centering\arraybackslash}p{0.7cm}|}{5} & & \\ \cline{1-8}
     
    1) Evaluation of the detection performances of the three UAD models  & $UAD_1$, $UAD_2$ and $UAD_3$ & & & & & \ding{51} & Figure \ref{fig:cluster_maps} for visual results \newline Table \ref{tab:detection_res} for performances details \\ \cline{1-8}

\end{tabularx}
\label{tab:exp_summary}
\end{table*}

\subsection{Generation of synthetic PET images from T1 healthy controls}

For both the 2.5D and 3D-patch approaches, we explore in total 4 variants of GANs for paired examples based on Simple-GAN or traditional Cycle-GAN architectures both with and without the proposed MSE extra loss term. 
We take ResNet as the backbone architecture of both generators with 9 residual blocks for the 2.5D approach and 2 residual blocks for the fully 3D-patch configuration. 
PatchGAN is selected for the discriminators following the architectures proposed in \cite{CycleGAN2017}.
All models were written by using PyTorch version 1.3.1 and we took python code provided by \cite{CycleGAN2017} as a baseline and available at \url{https://github.com/junyanz/pytorch-CycleGAN-and-pix2pix/}.
Detailed architectures and types of layers are depicted in Figure \ref{fig:ResNet} and Figure \ref{fig:PatchGAN} of Appendix \ref{sec:AppendixB} for the generator and discriminator, respectively. 

A 4-fold cross-validation performance study is conducted based on the 35 $DB_{C1}$ exams based on training parameters described in Appendix B. 
During the training, SSIM, as defined in section \ref{sec:OOD-metric}, between real and synthetic validation images serves as a quality metric to define the optimal configuration. It is computed on the generated triplets of 2D images for the 2.5 approach and on the generated patches for the 3D-patch approach and averaged over all validation control samples. SSIM is not computed on the 3D reconstructed and post-processed (Gaussian smoothing and histogram normalization) validation volumes at each epoch for time consideration.
The selected best 2.5D and 3D-patch GAN models from this cross-validation experiment are then fully retrained on the $DB_{C1}$ dataset considering 29 training, 3 validation and 3 test controls. 
Training parameters remain the same as for the 4-cross-fold validation experiment. The validation set is used to select the best model based on the SSIM metric. Training is stopped after 74 epochs for the 2.5 model and 24 epochs for the 3D-patch model.

The synthetic PET images are reconstructed following the method described in section \ref{sec:GAN_models}. A 3D Gaussian smoothing is applied on both 2.5D and 3D-patch reconstructed PET images to reduce border effects. Among a range of values between 0 and 3 mm FHWM, the value of 1.5 mm is shown to produce the best SSIM values. Finally, as detailed in \ref{sec:GAN_models}, histogram matching is applied as a final post-processing.

\subsection{Visual quality metrics assessment of the synthetic PET data}
\label{subsec:visual metrics}

The three visual quality metrics, namely PSNR, SSIM and LPIPS are computed on the validation images of each of the 4 folds of the cross-validation experiment.
Preliminary analysis of synthetic PET images generated by the 2.5D models demonstrated the added value of the MSE loss term on both qualitative and quantitative visual performance, so we chose not to evaluate the performance of the 3D-patch models trained without this MSE loss term.
For each of the evaluated models, we report the mean value over all images as well as the corresponding standard deviation.

\subsection{OOD performance metrics}

As stated in section \ref{sec:OOD-metric}, our purpose is to demonstrate that training of the autoencoder described in section \ref{sec:brain_ano_cad} with paired T1 MRI and synthetic PET (referred to as \textit{fake paired}) data produces \textit{image} and \textit{latent} space distributions similar to those achieved with the same auto-encoder trained on real paired T1 MRI and PET data. As a proxy experiment, we input true and fake pairs of MR and PET images to the auto-encoder trained on fake paired samples and evaluate if these test samples are ID or OOD ones, based on the two OOD metrics defined in section \ref{sec:OOD-metric}, namely, reconstruction error $MSE$ and Mahalanobis distance $D_m$. As the auto-encoder has been trained on fake paired data, the real PET and T1 MRI data of $DB_{C1}$ should appear as \textit{ID} if the fake PET images of $DB_{C2}^{2.5D}$  and $DB_{C2}^{3Dpatch}$ correctly mimic the true FDG distributions images, and may appear as \textit{OOD} if the fake PET images fail.

For this study, we consider the UAD detection model of Figure \ref{fig:brain_ano_CAD} whose architecture is described in the following section \ref{sec:brain_ano_cad}.
Details on the hyperparameters setting of this model are reported in Appendix \ref{sec:AppendixC}.
As shown on Table \ref{tab:exp_summary}, this auto-encoder is trained separately on the $DB_{C2}^{2.5D}$ and $DB_{C2}^{3Dpatch}$ fake paired datasets, to generate the $UAD_2$ and $UAD_3$ models, respectively.
For each model, we compute the $MSE$ and $D_m$ metrics on the 35 paired true MR T1 and PET exams of $DB_{C1}$, as well as on the 17 epilepsy patients of $DB_{ep}$ and on the 5 test control samples of $DB_{C2}^{2.5D}$ (for $UAD_2$) or $DB_{C2}^{3Dpatch}$ (for $UAD_3$).

\subsection{Application of synthetic PET data to the training of a brain unsupervised anomaly detection model for epilepsy patients screening}

In this section, we evaluate detection performance of the three brain UAD models described in previous section \ref{sec:brain_ano_cad} and Table \ref{tab:exp_summary}, namely $UAD_1$ trained on $DB_{C1}$, $UAD_2$ trained on $DB_{C2}^{2.5D}$ and $UAD_3$ trained $DB_{C2}^{3Dpatch}$.

The trained $UAD_1$, $UAD_2$ and $UAD_3$ models are tested on the 17 patients of $DB_{ep}$ with confirmed medically intractable EZ localisations as well as on patient $R$ with surgical resection of the right temporal lobe (see Table \ref{tab:detection_res}). Note that these 17 epilepsy patients correspond to difficult detection cases. Their T1 and FDG PET exams are indeed considered as normal, meaning that the hypometabolic lesions are subtle and barely visible by expert neurologists.
For each patient of $DB_{ep}$, we report the ranking of the detected cluster (among 10 candidates) that eventually intersects the ground truth annotation, as well as the overall detection rate and mean rank of the detected clusters averaged over the 17 epilepsy patients (excluding patient $R$). We also perform a qualitative visual analysis of the detected cluster maps.

For all experiments, we take the code from the model proposed in \cite{alaverdyan_MEDIA2020} and available at \url{https://github.com/clartizien/Brain-AnoCAD-v2021/}. As mentioned above, Appendix \ref{sec:AppendixC} provides details on the hyperparameters setting of the UAD model and post-processing of the anomaly score maps, as well as on the performance analysis. 

\section{Results}

\subsection{Visual performance metrics}

Table \ref{tab:Quantitative_results} reports mean SSIM, PSNR and LPIPS metrics with corresponding standard deviations computed over all reconstructed and post-processed (Gaussian smoothing and histogram normalization) 3D validation volumes and all folds for all considered models.
The 3D-patch Cycle-GAN with MSE loss is shown to perform the best among the 6 models considered in this study. 
Two-tailed Wilcoxon signed rank tests yield no significant differences between the 2.5D and 3D-patch Cycle-GAN models with MSE loss for the PSNR metric (p-value = $0.79$) while p-values of $4$x$10^{-4}$ and $8.6$x$10^{-8}$ are achieved for the LPIPS and SSIM metrics, respectively, in favor of the 3D-patch method.
Also, note that our proposition to add the MSE loss term to the global loss of the 2.5D Cycle-GAN model allows a significant improvement of all three metrics with p-values of $5.8$x$10^{-11}$, $3.8$x$10^{-7}$ and $1.7$x$10^{-10}$ for the SSIM, PSNR and LPIPS metrics, respectively. A significant improvement of these metrics is also observed for the Simple-GAN model with p-values of $5.8$x$10^{-11}$, $4.1$x$10^{-9}$ and $5.8$x$10^{-11}$ for the SSIM, PSNR and LPIPS metrics, respectively. 

\begin{table*}[]
\caption{\label{tab:test_reslts}Average visual quality metrics computed on synthetic PET exams generated from T1 MRI of 35 healthy subjects of $DB_{C1}$. The best model in each category (2.5D or 3D-patch) is in bold.}
\begin{tabular}{|l|l|l|l|l|}
\hline
\textbf{Configuration}    & \textbf{Model}           & \textbf{SSIM} & \textbf{PSNR} & \textbf{LPIPS} \\ \hline
\multirow{4}{*}{2.5D}  & Simple-GAN               & $0.825\pm0.020$    & $21.489\pm0.890$   & $0.033\pm0.006$    \\ \cline{2-5} 
                          & Simple-GAN with MSE loss & $0.880\pm0.020$    & $23.542\pm1.430$   & $0.022\pm0.006$   \\ \cline{2-5} 
                          & Cycle-GAN                & $0.884\pm0.020$   & $23.700\pm1.430$    & $0.023\pm0.006$     \\ \cline{2-5} 
                          & \textbf{Cycle-GAN with MSE loss}  & $\textbf{0.886}\pm\textbf{0.019}$    & $\textbf{23.742}\pm\textbf{1.260}$   & $\textbf{0.021}\pm\textbf{0.004}$    \\ \hline
\multirow{2}{*}{3D-patch} & Simple-GAN with MSE loss & $0.883\pm0.020$   & $23.100\pm1.380$  & $0.021\pm0.004$    \\ \cline{2-5} 
                          & \textbf{Cycle-GAN with MSE loss}  & $\textbf{0.897}\pm\textbf{0.019}$   & $\textbf{23.820}\pm\textbf{1.720}$    & $\textbf{0.019}\pm\textbf{0.005}$    \\ \hline
\end{tabular}
\label{tab:Quantitative_results}
\end{table*}

\begin{figure*}[]
\centering
\includegraphics[width=12cm]{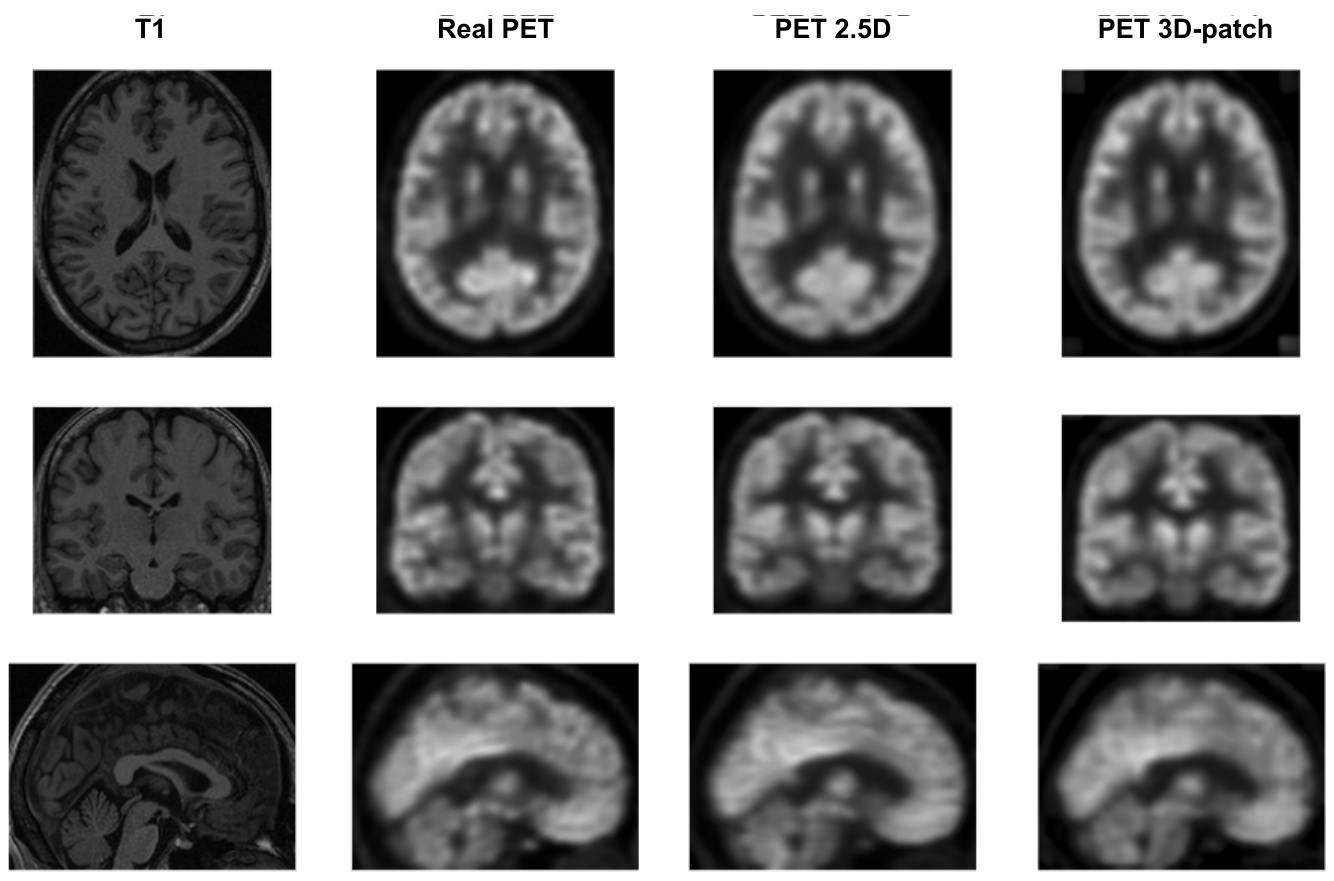}
\caption{Qualitative result on one validation control subject of $DB_{C1}$. First left column: original T1 MRI, then, from left to right:  original FDG PET image, synthetic PET images generated with 2.5D and 3D-patch Cycle-GAN models with MSE loss.}
\label{fig:SyntheticPET}
\vspace{-3mm}
\end{figure*}

In the following, we consider the best performing of each configuration, namely 2.5D and 3D-patch Cycle-GAN models with MSE loss. Example synthetic PET data generated by these two configurations of Cycle-GAN models from the same T1 MRI of a test control subject of $DB_{C1}$ are illustrated in Figure \ref{fig:SyntheticPET} and compared with the reference PET image of this subject. Both models allow for generating visually realistic FDG PET data.

\subsection{Task-oriented performance metrics}

\begin{figure*}
     \centering
     \begin{subfigure}[b]{0.45\textwidth}
         \centering
         \includegraphics[width=\textwidth]{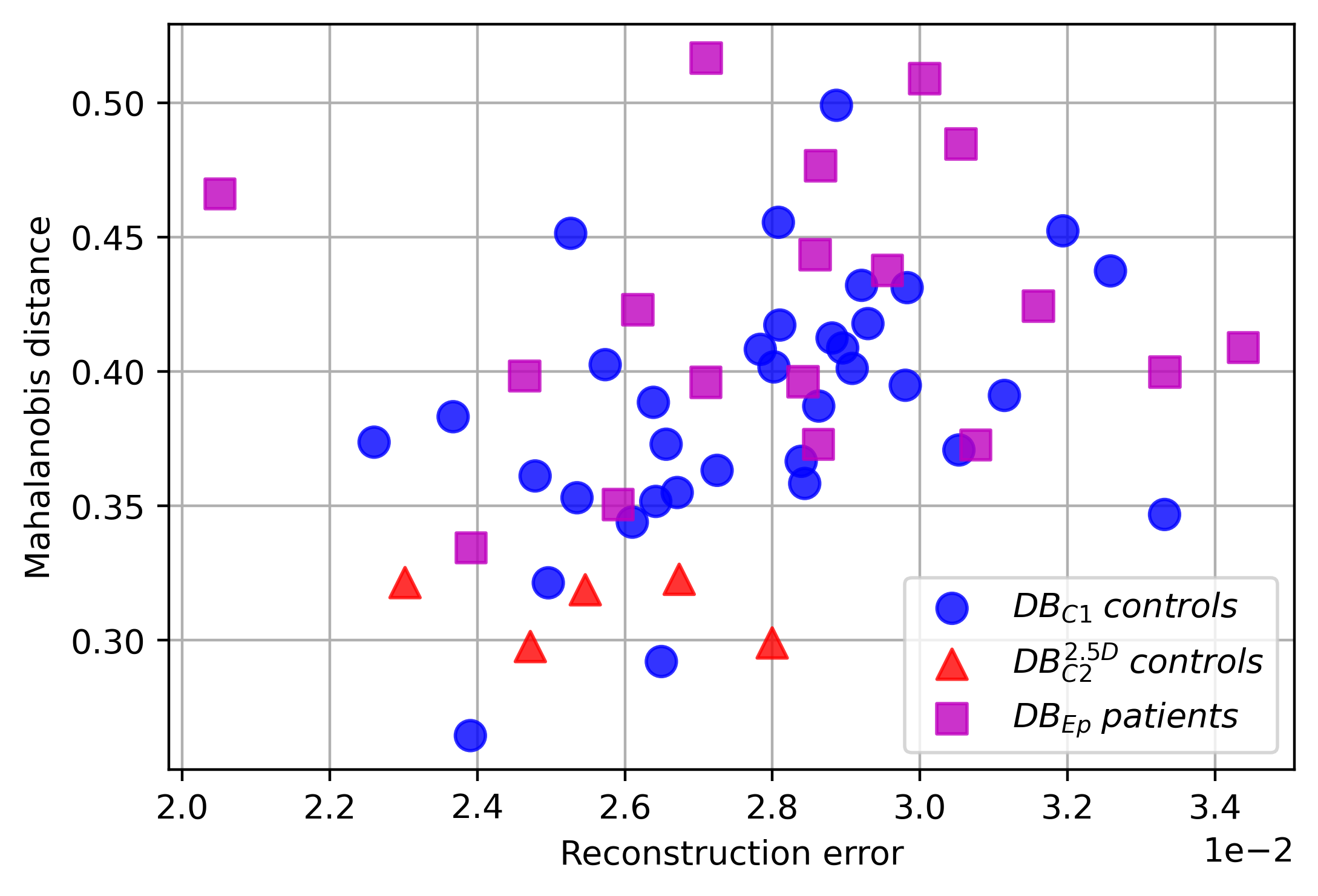}
         \caption{}
         \label{subfig;2.5D_scatter_plot}
     \end{subfigure}
     \begin{subfigure}[b]{0.45\textwidth}
         \centering
         \includegraphics[width=\textwidth]{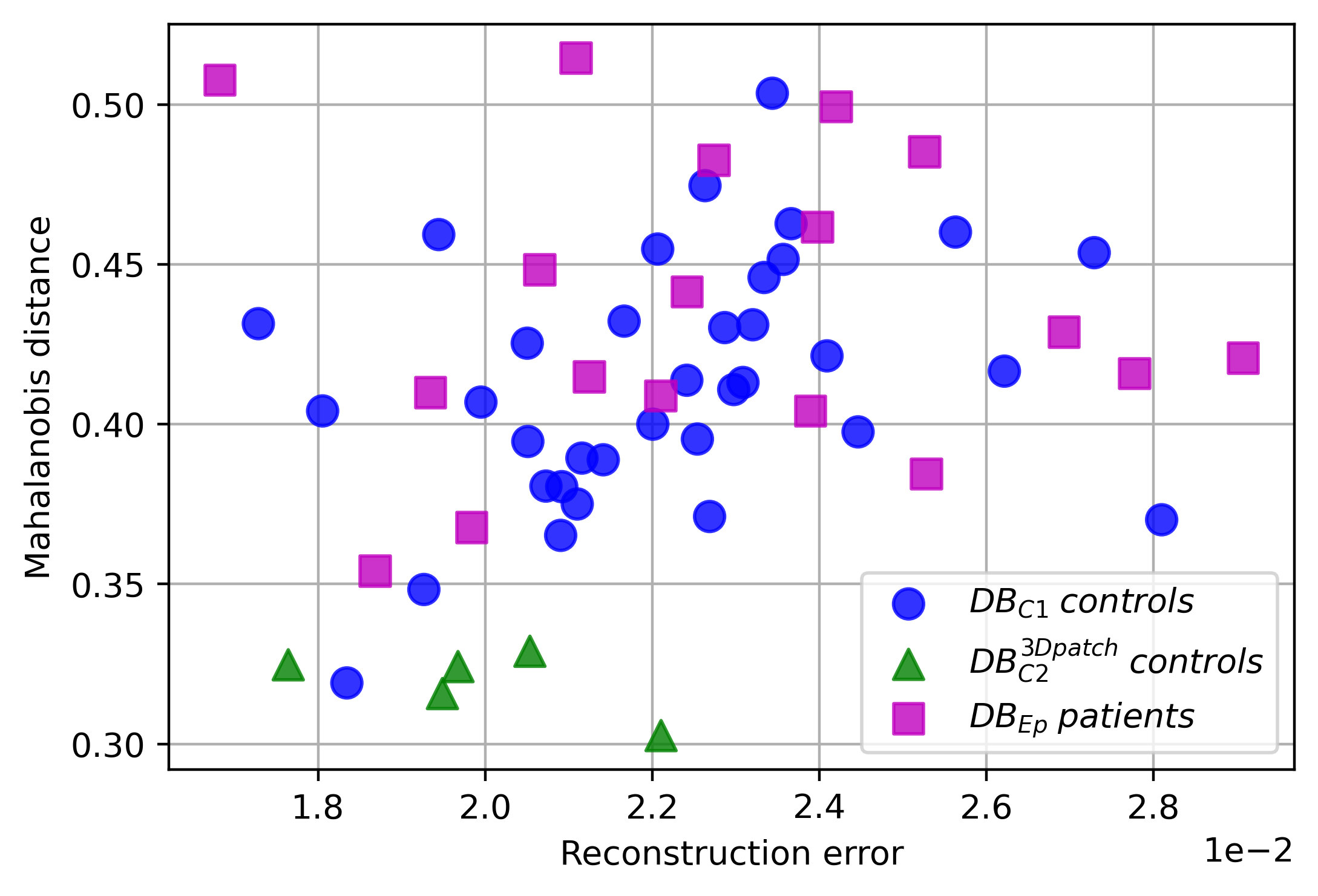}
         \caption{}
         \label{subfig:3Dpatch_scatter_plot}
     \end{subfigure}
     \hfill
    \caption{Mahalanobis distance $D_m$ and reconstruction error $MSE$ on test subjects inputted to the $UAD$ model of \cite{alaverdyan_MEDIA2020}. Blue points correspond to the 35 real controls from $DB_{C1}$, purple squares to the 18 patients of $DB_{ep}$. a) Results for $UAD_2$ trained on $DB_{C2}^{2.5D}$, red triangles correspond to 5 test control samples of $DB_{C2}^{2.5D}$. b) Results for $UAD_3$ trained on $DB_{C2}^{3Dpatch}$, green triangles correspond to 5 test control samples of $DB_{C2}^{3Dpatch}$.}
    \label{fig:2.5D_3Dpatch_scatter_plot}
\end{figure*}

Figure \ref{fig:2.5D_3Dpatch_scatter_plot} shows the relationship between the mean normalized Mahalanobis distance $D_m$ and the mean reconstruction error $MSE$ estimated on test subjects from the $UAD_2$ and $UAD_3$ models trained on 35 samples of $DB_{C2}^{2.5D}$ and 35 samples of $DB_{C2}^{3Dpatch}$, respectively. The blue circle symbols and the purple square ones report metrics computed on the 3D real normal samples of $DB_{C1}$ and 18 patients of $DB_{ep}$, respectively, while the red and green triangles correspond to the 5 remaining test control samples of $DB_{C2}^{2.5D}$ and $DB_{C2}^{3Dpatch}$, respectively.

For all controls and patients, we observe the same order of reconstruction error with average values of $0.027\pm0.002$, $0.025\pm0.002$, $0.028\pm0.003$ for the 35 $DB_{C1}$ controls (blue circle), 5 $DB_{C2}^{2.5D}$ test controls (red triangle) and 18 $DB_{ep}$ patients (purple square) images reconstructed from the 2.5D Cycle-GAN model ($UAD_2$). Average MSE for images reconstructed from the $UAD_3$ model are $0.022\pm0.002$, $0.019\pm0.002$ and $0.022\pm0.003$ for the 35 $DB_{C1}$ controls (blue circles), 5 $DB_{C2}^{3Dpatch}$ controls (green triangle) and 18 $DB_{ep}$ patients (purple square), respectively. The difference is more noticeable in the values of the Mahalanobis distance computed in the latent space with average values of $0.39\pm0.04$ ($0.41\pm0.04$) for the 35 $DB_{C1}$ controls (blue circle), $0.42\pm 0.05$ ($0.43\pm 0.04$) for the 18 $DB_{ep}$ patients (purple square) and $0.31\pm0.01$ ($0.32\pm0.01$) for 5 $DB_{C2}$ (red ($DB_{C2}^{2.5D}$) and green ($DB_{C2}^{3Dpatch}$) triangles) latent representations learned from the $UAD_2$ ($UAD_3$) model.
These results suggest that the distributions of $DB_{C1}$, $DB_{C2}^{2.5D}$ and $DB_{C2}^{3Dpatch}$ control samples reasonably overlap in the image and latent spaces thus meaning that fake paired images of $DB_{C2}^{2.5D}$  and $DB_{C2}^{3Dpatch}$ can be considered as inliers of the distribution of true control samples of $DB_{C1}$. Also note that the patient and control scatter plots are well-mixed, thus meaning that the patients can not be considered outliers of the distribution of control samples. This underlines the difficulty of the detection task considered in this study, where epilepsy lesions are so subtle that they do not impact the global patient-level OOD metrics considered in this study. These results are further discussed in section \ref{sec:discussion}.

\subsection{Impact of synthetic PET data on brain anomaly detection performance}
\label{sec:AD_perf}

Detection performance results are reported in Table \ref{tab:detection_res}. We remind (see section \ref{sec:data}) that patient $R$ was added to evaluate the ability of the UAD models to detect \textit{non-subtle anomalies}, here a surgical resection localization in the right temporal pole. Detection performance achieved for this patient is not included in the overall metrics reported on the last two lines of Table \ref{tab:detection_res}, thus aggregating performance estimated on the 19 small and subtle epileptogenic lesions of patients $A$ to $Q$. The best-reported detection performance was achieved with the $UAD_3$ model trained on $DB_{C2}^{3Dpatch}$, with $74\%$ sensitivity (14 out of the 19 lesions) and a mean reported rank of $2.1$, meaning that the detected epilepsy lesions were, on average, among the top 3 most suspicious clusters reported by this model. The $UAD_2$ model trained on $DB_{C2}^{2.5D}$ achieved $58\%$ sensitivity (11 out of the 19 lesions) and a mean reported rank of $2.4$, thus outperforming the model trained on the real paired T1 MRI and FDG PET of $DB_{C1}^{3Dpatch}$ whose sensitivity and mean rank were of $42\%$ (8 out of 19 lesions) and 3.9, respectively. These results are on par with those of the quantitative visual analysis reported in Table \ref{tab:Quantitative_results} and show that images obtained from the 3D-patch CycleGAN model appeared to be visually realistic as well as adapted to the training of the UAD model. Further analysis of the comparative performance achieved by these different models, in particular the somewhat counter-intuitive result that the UAD models trained on paired fake PET data outperform that trained on true paired data, is discussed in section \ref{sec:discussion}.

\begin{figure*}[h!]
\centering
\includegraphics[width=12cm]{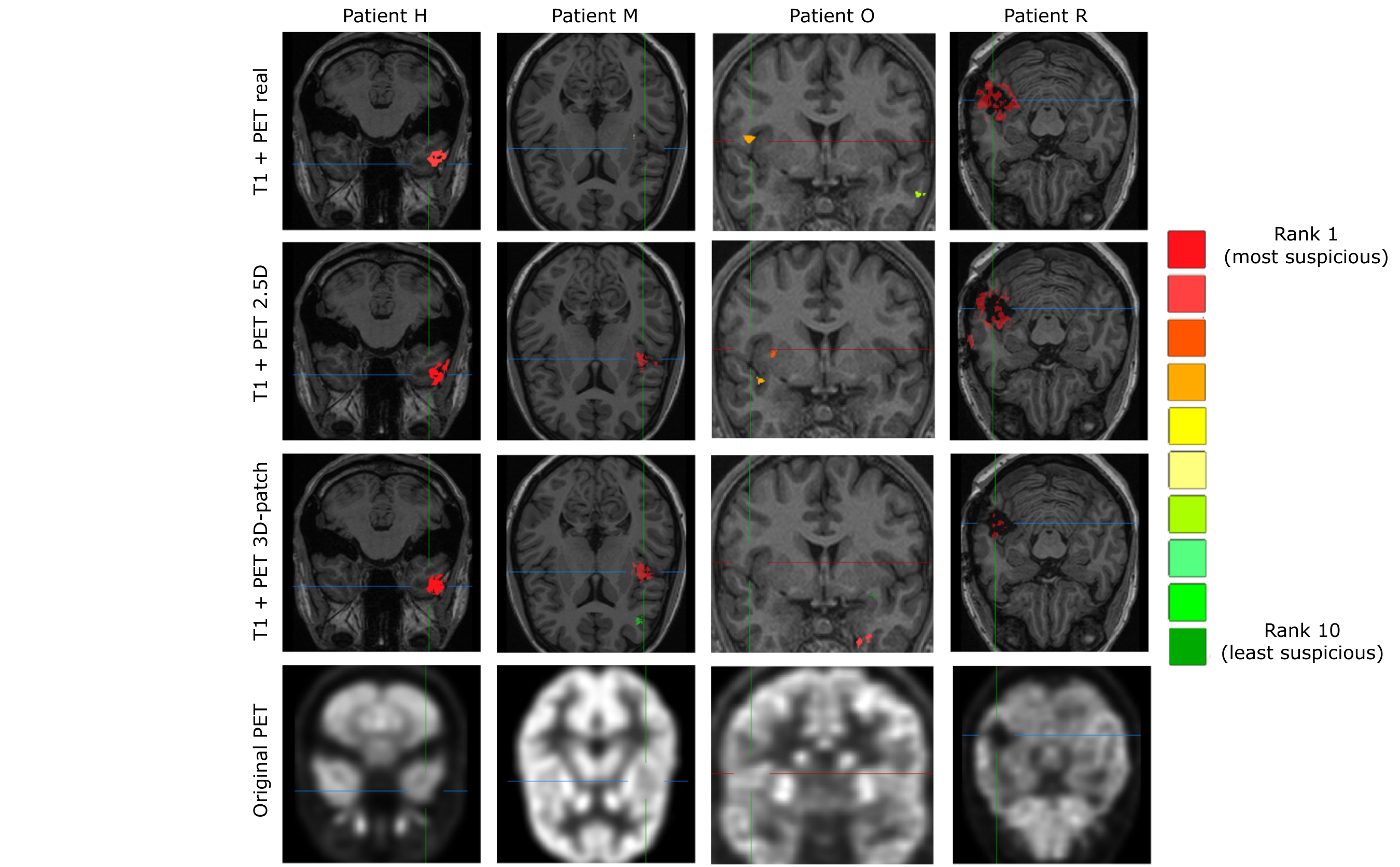}
\caption{Example cluster maps overlaid on T1 MRI  by the detection models, from top to bottom: $UAD_1$, $UAD_2$ and $UAD_3$, respectively. Selected transverse or coronal slices of patients $H$, $M$, and $O$ are centered on confirmed EZ localisations in various areas of the brain, namely the left temporal pole, left insula, and right precentral gyrus. Illustration of patient $R$ depicts performance for the detection of a large surgical resection area located in the right temporal lobe.
The cursor points to suspicious anatomical regions. The color bar displays the most suspicious cluster of rank 1 as bright red and the least suspicious detected cluster of rank 10 as dark green.}
\label{fig:cluster_maps}
\vspace{-5mm}
\end{figure*}

Figure \ref{fig:cluster_maps} illustrates anomaly cluster maps derived from the three detection models on 4 test epilepsy patients. These visual results confirm the quantitative performance reported in Table \ref{tab:detection_res}. Lesion of Patient $H$ in the left temporal pole was correctly detected by all models but with the highest rank obtained by the $UAD_3$ model trained on T1 and synthetic PET data generated by the 3D-patch cycleGAN. Lesion of Patient $M$ located in the left insula was correctly detected by the $UAD_3$  and $UAD_2$ model with the highest suspicious rank, but it was missed by the $UAD_1$ model trained on real T1 and PET data. Lesion of Patient $O$ located in the vicinity of the right precentral gyrus and operculum was correctly detected with $UAD_1$ and $UAD_2$ models with ranks of 4 and 3, respectively, but it was missed by $UAD_3$ model. At the most right, the surgical resection zone of Patient $R$ in the right temporal lobe was correctly detected by all three models with high confidence.

\begin{table*}
\caption{Performance of the brain anomaly detection model trained on the three databases : from left to right : $UAD_1$ : 35 real T1 and PET samples of $DB_{C1}$, $UAD_2$ : 35 paired true T1 MR and fake PET of $DB_{C2}^{2.5D}$, $UAD_3$ : 35 paired true T1 MR and fake PET of $DB_{C2}^{3Dpatch}$. {\ding{51}} denotes a true detection followed by its rank inside parentheses. {\ding{55}} denotes no true positive detection meaning that the lesion was not detected among the 10 highest ranked clusters detected by the model. Lines 3 and 2 starting from the bottom report the total number of detected lesions over all epiletogenic patients (Patient $A$ to $R$) as well as the mean rank score assigned by each model. L : Left, R : Right}
\def\arraystretch{1.5}
\begin{tabular}{|l|l|c|c|c|c|}
\hline
\textbf{Patient} & \textbf{EZ location} &  \begin{tabular}[c]{@{}c@{}}\textbf{T1+PET real}\\$UAD_1$\end{tabular}& \begin{tabular}[c]{@{}c@{}}\textbf{T1+PET 2.5D}\\$UAD_2$\end{tabular} & \begin{tabular}[c]{@{}c@{}}\textbf{T1+PET 3D-patch}\\$UAD_3$\end{tabular}  \\
\hline

Patient $A$   & Temporal Lobe L    &  \ding{55}       & \ding{55}        & \ding{55}       \\ \hline
Patient $B$   & Temporal Lobe L    &  \ding{51} (5)   & \ding{51} (4)    & \ding{51} (1)   \\ \hline
Patient $C$   & Temporal Lobe R    &   \ding{51} (3)       &  \ding{51} (4)        &  \ding{51} (2)       \\ \hline
Patient $D$   & Temporal Pole R    &  \ding{51} (8)   & \ding{55}        & \ding{51} (1)   \\ \hline
Patient $E$   & Temporal Pole R    &  \ding{55}       & \ding{55}        & \ding{55}       \\ \hline
Patient $F$   & Temporal Pole R    &  \ding{55}       & \ding{55}        & \ding{51} (1)   \\ \hline
Patient $G$   & Temporal Pole R    &  \ding{55}       & \ding{51} (2)    & \ding{51} (1)  \\ \hline
\multirow{2}{*}{Patient $H$}   & Temporal Pole L    &  \ding{51} (2)   & \ding{51} (3)    & \ding{51} (1)  \\ \cline{2-5}
                               & Insula L             &  \ding{55}       & \ding{55}        & \ding{51} (5)  \\ \hline
\multirow{2}{*}{Patient $I$}   & Insula R             &  \ding{55}       & \ding{51} (4)    & \ding{51} (5)  \\ \cline{2-5}
                               & Temporal Pole R    &  \ding{55}       & \ding{55}        & \ding{55}      \\ \hline
Patient $J$   & Insula L             &  \ding{51} (1)   & \ding{51} (1)    & \ding{51} (3)   \\ \hline
Patient $K$   & Insula R             &  \ding{55}       & \ding{51} (3)    & \ding{51} (5)   \\ \hline
Patient $L$   & Insula R            &  \ding{51} (7)   & \ding{51} (1)    & \ding{51} (1)   \\ \hline
Patient $M$   & Insula L           &  \ding{55}       & \ding{51} (1)    & \ding{51} (1)   \\ \hline
Patient $N$   & Insula R            &  \ding{51} (1)   & \ding{51} (1)    & \ding{51} (1)   \\ \hline
Patient $O$   & Other              &  \ding{51} (4)   & \ding{51} (3)    & \ding{55}       \\ \hline
Patient $P$   & Other              &  \ding{55}       & \ding{55}        & \ding{55}       \\ \hline
Patient $Q$   & Other              &  \ding{55}       & \ding{55}        & \ding{51} (2)   \\ \hline
\multicolumn{2}{|l|} {\textbf{Overall \# of lesion detections (19 max)}} & \textbf{8} & \textbf{11} & \textbf{14}  \\ \hline
\multicolumn{2}{|l|} {\textbf{Mean Rank}} & \textbf{3.9} & \textbf{2.4} & \textbf{2.1}  \\
\hline \hline
Patient $R$   & surgical resection R    &  \ding{51} (1)   & \ding{51} (1)    & \ding{51} (1)   \\ \hline
\end{tabular}
\label{tab:detection_res}
\end{table*}

\section{Discussion}
\label{sec:discussion}

In this study, we demonstrate that realistic FDG PET exams of healthy subjects can be generated from GAN-based architectures with T1 MRI as input. This is assessed by the quantitative visual metrics, such as PSNR and SSIM reported in Table \ref{tab:Quantitative_results} as well as qualitatively, based on example samples reported in Figure \ref{fig:SyntheticPET}. Quantitative results show that our proposal to train the standard CycleGAN architecture with paired T1 and PET data and additional MSE loss term allowed notable improvements in comparison to standard CycleGAN.
We also show that these synthetic PET data could efficiently replace true FDG PET images into multi-modality normative databases to train unsupervised anomaly detection models. This was first assessed based on the OOD metrics reported in Figure \ref{fig:2.5D_3Dpatch_scatter_plot} indicating a reasonable overlap of the metrics computed on the paired T1 and real PET and paired T1 and synthetic PET data. This study provides clues of evidence that an UAD detection model trained with normative fake PET would not silently fail when tested on true patient PET data. The detection performance analysis conducted in section \ref{sec:AD_perf} allows confirming conclusions drawn from the OOD metrics. 

As stated in the Introduction, this paper significantly extends our preliminary work in \cite{Zotova_Sashimi21}. We improved the visual quality of synthetic PET data by performing histogram matching of the generated synthetic PET, thus producing synthetic PET images that closely match the original ones. 
We also introduce diagnostic task-oriented quality metrics of the synthetic imaging data and strengthen the use case application focusing on the detection of epilepsy lesions in paired FDG PET and T1 MR data. We included an extensive performance evaluation of the proposed UAD model on a homogeneous and extended series of clinical epilepsy patients with confirmed lesion localization and surgery outcome as well as paired T1 MRI and PET exams realized in similar conditions. 

Results reported in Table \ref{tab:detection_res} demonstrate that UAD models trained on synthetic PET data allow achieving detection performance on par with that achieved with the same model trained on real PET data. 
The slight improvement of detection performance observed with the models trained on synthetic paired data might be explained by the lower mean and standard error values of the Mahalanobis distance $D_m$ estimated on the inliers fake pairs, respectively the 5 test samples of $DB_{C2}^{2.5D}$ (red triangles) in Figure \ref{subfig;2.5D_scatter_plot} and the 5 test samples of and $DB_{C2}^{3Dpatch}$ (green triangles) in Figure \ref{subfig:3Dpatch_scatter_plot}. This might indicate that the latent distribution of the synthetic representation $z$ is denser than that achieved with the real paired T1 MR and FDG PET samples. The denser the normative distribution in the latent representation space, the more compact the density support estimated by the OC-SVM model, and the more sensitive it is to any deviation from normality at inference. One possible explanation for the assumed higher density of the synthetic latent distribution is that the synthetic PET data have lower inter-individual pattern variability than the real population. In summary, the $UAD_2$ and $UAD_3$ models trained on paired synthetic data might be more sensitive to any subtle deviation from the normative pattern, including those originating from subtle epilepsy lesions. This attempted explanation should be interpreted with caution. Further investigation including reproducibility analysis based on an extended database, is required to confirm this trend. 

Although the main focus of this study was not to report on the UAD detection performance, it is still worth adding some comparative analysis with that reported in the literature.
Our model trained on paired T1 MR and synthetic 3D-patch PET data achieved sensitivity around $74\%$ while we reported sensitivity around $60\%$ with the same UAD model trained on real T1 MR and FLAIR data in \cite{alaverdyan_MEDIA2020}. This highlights the added value of the PET modality for this challenging detection task. 
The best-achieved detection sensitivity of 74\% may seem low. This value, however, has to be compared with the very low sensitivity of human experts on these difficult diagnostic cases and is on par with reported values in the literature.
Indeed, visual performance reported in this study is in the range of values obtained in \cite{yaakub2019pseudo} and \cite{Flaus_EpilepsiaOpen23} for the same task of generating FDG PET images of control subjects based on their T1 MRI. Both studies implemented GAN-based architectures with resulting PSNR values of 23.2 $\pm$ 2.3 in \cite{yaakub2019pseudo} and  35$\pm$ 3.8 in \cite{Flaus_EpilepsiaOpen23}. As stated in section \ref{sec:related_work}, they also used synthetic data to detect epileptogenic lesions, however in a slightly different paradigm than ours. Indeed, their detection task was based on PET images only. It consisted of synthesizing pseudo-healthy PET data from T1 MRI of epileptic patients and deriving a z-score by subtracting the patient's true PET from their estimated pseudo-normal. Thresholding this z-score map allowed the outlining of suspicious areas. This detection method assumes that the generative synthesis model, trained on paired T1 and FDG PET of control healthy subjects only, will not introduce pathological FDG uptake in the generated synthetic PET image, at inference, when inputted a pathological T1 MR exam. Flaus et al report sensitivity of $67\%$ in 12 MRI-positive versus $50\%$ in 8 MRI-negative patients \cite{Flaus_EpilepsiaOpen23},  while Yaakub et al report $92\%$ sensitivity in 7 MRI-positive versus $75\%$ in 13 MRI-negative exams \cite{yaakub2019pseudo}.

This study builds on the CycleGAN model that demonstrates impressive performance in translating T1 MRI to FDG PET imaging, as reported in section \ref{sec:related_work}. We improved the visual quality of the synthetic PET images by training this architecture with paired FDG and T1 MRI exams and adding the MSE loss term in eq. \ref{Eq_MSEloss}. The quantitative visual metrics that we achieved are in par with state-of-the art performance of GAN-based models recently reported in the review paper of \cite{Dayarathna_MEDIA24}. This review also emphasizes that such adversarial models are the best suited for this type of cross-modal translation, outperforming both transformer or diffusion models which are shown best adapted for other image synthesis tasks. Transformers indeed allow unifying various source-target modality configurations into a single model, which means taking any subset of input contrasts or modality and synthesizing those that are missing as in a unified and unique framework \cite{Dalmaz_TMI22, Liu_TMI23}. Diffusion models are shown very efficient in generating high quality synthetic images conditioned to text, clinical variables, as well as to other imaging modalities as in \cite{xie_arxiv24}, which is different from the pursued objective of this study. Despite these arguments, the lack of comparison with other image-to-image translation models, such as transformer or diffusion models may limit the scope of our study.

Another limitation of our work is the limited size of our patient cohort. Conclusions regarding both the OOD evaluation and detection performance of the UAD model should be confirmed on a larger cohort.

Also, our patient cohort was homogeneous, meaning that all MRI T1 and FDG PET exams were acquired on the same scanner type as the one that served to generate the training normative dataset of the generative CycleGAN model. We thus need to validate that the UAD model would generalize well with patient data acquired on different scanners. The OOD metrics that we proposed are easy and fast to compute metrics enabling to evaluate if the patient data are ID or OOD for the considered UAD model. We could thus project any new patient in the plots of Figure \ref{fig:2.5D_3Dpatch_scatter_plot} to quantify its distance to the ID normative distribution. In our preliminary study \cite{Zotova_Sashimi21}, where we trained one UAD model on synthetic FDG PET data only, we could show that the model was quite strong in detecting EZ in patient FDG PET exams that were acquired on different scanner and with different noise patterns. 

One perspective to this work would be to assess the reproducibility of our results based on an extended validation study with more control subjects and patients. 
We also plan to further assess the added value of attention modules, e.g. based on the transformer models that are shown to perform well for the imputation of missing imaging modalities. 
Perspectives to improve detection performance with heterogeneous patient data would be to mimic this variability in the generation of synthetic PET normative data to increase robustness of the brain anomaly detection model.
Finally, as recently investigated in \cite{Pan_MICCAI21} and \cite{Pan_TPAMI22}, one promising way is to design models that jointly tackle the synthesis of the modality and the diagnostic task at hand, by coupling synthesis and diagnostic (e.g. classification) networks.

\section{Conclusion}
This study demonstrates that realistic FDG PET exams of healthy subjects can be generated from GAN-based architectures with T1 MRI as input.
But our main claim is to show that these synthetic normative PET data could efficiently replace true FDG PET images into multi-modality normative databases to train unsupervised anomaly detection models. This is first assessed based on tailored out-of-the distribution metrics and on a detection performance analysis conducted on 17 real paired T1 MRI and FDG PET exams of epilepsy patients. 

\section{Acknowledgments}

This work was granted access to the HPC resources of IDRIS under the allocation 2021-AD011011938 made by GENCI. It received funds from the Région Auvergne-Rhône-Alpes through the TADALOT project and from the French National Research Agency (ANR) through the IMAGINA project (ANR-18-CE17-0012). It was performed within the framework of the SEIZURE project (ANR-24-CE45-4399-01) and LABEX PRIMES (ANR-11-LABX-0063) of Université de Lyon operated by ANR.

\printcredits

\appendix

\section{Visual quality metrics}
\label{sec:AppendixA}

Mean squared error (MSE) estimates the mean pixel-based error between the generated ($x$) and ground truth ($y$) images as
\begin{equation}
\label{MSE_eq}
\ MSE=\frac{1}{mn}\sum_{i=0}^{m-1}\sum_{j=0}^{n-1}\left| x(i,j)-y(i,j) \right|^{2}
\end{equation}
for images of dimension $m \times n$. \\

Peak Signal-to-Noise Ratio (PSNR) computes the ratio between the maximum possible power (pixel intensity value) of the generated image $y'$ and the MSE defined in eq (\ref{MSE_eq}) characterizing the power of corrupting noise that affects the fidelity of its representation, as
\begin{equation}
\label{PSNR_eq}
\ PSNR = 20log_{10}\left( \frac{max_{x}}{\sqrt{MSE}} \right)
\end{equation}

\ where $max_{x}$ is the maximum possible pixel value of image $x$.\\

Structural similarity index metric (SSIM) \cite{wang2004image} estimates the perceptual difference between two images using the mean ($\mu$) and standard deviation ($\sigma$) over pixel values of the generated ($x$) and ground truth ($y$) images. Two constants, $c_1 = (0.01 L)^{2}$ and $c_2 = (0.03 L)^{2}$, are included to stabilize the division with low denominator, with L being the dynamic range of the pixel values, as

\begin{equation}
\label{SSIM_eq}
\ SSIM(x, y) = \frac{(2\mu_{x}\mu_y + c_1)\cdot(2\sigma_{xy}+c_2)}{(\mu^{2}_{x}+\mu^{2}_y+c_1)\cdot (\sigma^{2}_{x}+\sigma^{2}_y+c_2)}
\end{equation}

Learned Perceptual Image Patch Similarity (LPIPS)\cite{zhang2018unreasonable}  is used to judge the perceptual similarity between two images and has been shown to match human perception. 
It computes the distance between the activation maps of the generated $x$ and ground truth $y$ images from a pre-defined network.

\begin{equation}
\label{lpips}
d(y, x) = \sum_{l}^{} \frac{1}{H_lW_l}\sum_{hw}^{}\left\| w_l\odot(\widehat{y}_{hw}^l-\widehat{x}_{hw}^l) ) \right\|_2^2
\end{equation}

Where $\widehat{y}_{hw}^l$ is the extracted feature for layer $l$ normalized in the channel dimension and $w_l$ is a learned linear weight vector. As a pre-defined network, we used AlexNet. 

\section{Architecture details of the generator and discriminator backbones of the CycleGAN model}
\label{sec:AppendixB}

\begin{figure*}[h]
    \centering
    \includegraphics[width=16cm]{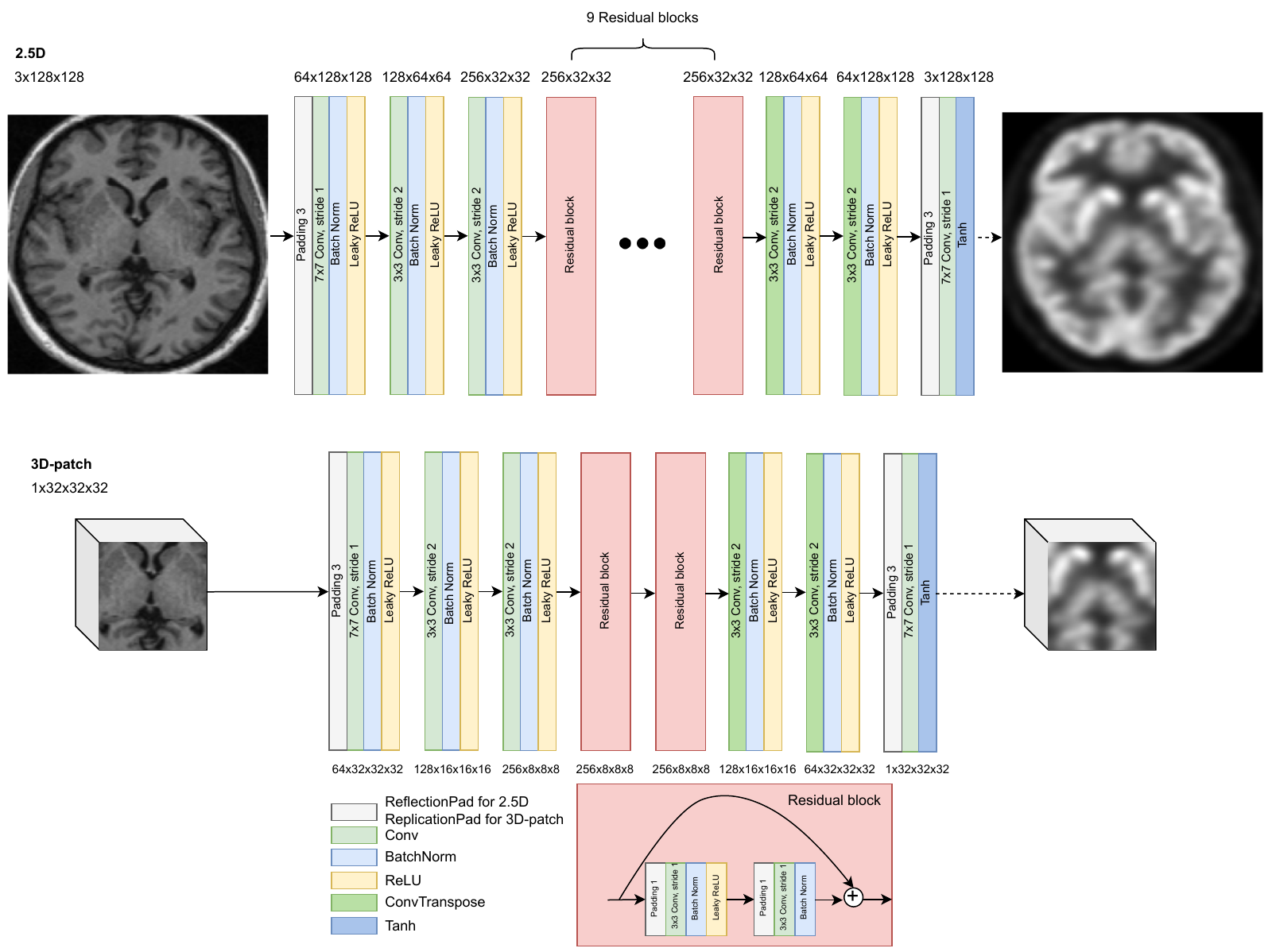}
    \caption{\label{fig:ResNet}ResNet architecture for 2.5D and 3D-patch generators
    }
\end{figure*} 

\begin{figure*}[h]
    \centering
    \includegraphics[width=10cm]{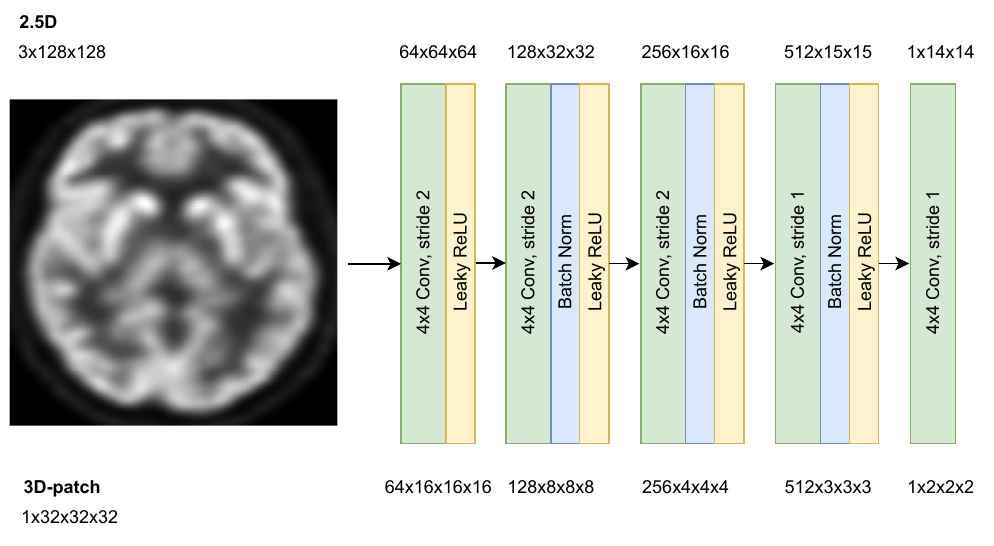}
    \caption{\label{fig:PatchGAN}PatchGAN architecture for 2.5D and 3D-patch discriminators}
\end{figure*}

Implementation details of the GANs training for each type of input data are as follows: \\

For the 2.5D approach:
\begin{itemize}
    \item 46 non-overlapping triplets of adjacent transverse slices are extracted per control training sample (covering the entire 3D volume) thus resulting in around 1200 training samples for each model.
    \item The models are trained for a maximum of 200 epochs with a batch size of 5 and Adam optimizer. 
    \item learning rate of 0.0002 is kept constant up to 100 epochs and linearly decayed to zero over the next 100 epochs.
\end{itemize}

For the 3D patch approach:
\begin{itemize}
    \item 6 069 mini-patches of size 32x32x32 are extracted for each control training subject with a stride of 8, thus leading to more than 160,000 training mini-volumes.
    \item The models are trained for a maximum of 100 epochs with a batch size of 10 respectively and Adam optimizer.
    \item The learning rate of 0.0002 is kept constant.
\end{itemize}

\section{Training of the UAD model, post-processing and performance analysis}
\label{sec:AppendixC}

In this work, we reproduce the 2D siamese architecture depicted in Figure \ref{fig:ocsvm} of \cite{alaverdyan_MEDIA2020} which inputs 15x15 patch size and results in a representation vector $z$ of dimension 64.
Images are scaled between 0 and 1 at the image level after removing top 1\% intensities. We extract 25 000 15x15 paired MRI and FDG PET patches per 3D image of each training control of $DB_{C2}$ to train the sAE network. Once the sAE is trained, we build one oc-SVM model per voxel with RBF kernel based on the latent representation vectors $z$ extracted for each voxel coordinate of the siamese autoencoder. 

During the test phase, each voxel of a given patient is matched against the corresponding oc-SVM classifier and is assigned the signed score output by the classifier. This yields a distance map for the given patient, which is then thresholded at some pre-chosen value and a 26-connectivity rule is applied to identify the connected components. These components are referred to as clusters.  
These clusters are ranked according to a criterion defined in \cite{alaverdyan_MEDIA2020}, which accounts for the cluster size and its average score. Such a ranking favors large clusters with the most negative average score. Using this ranking, we keep the top $n$ detections and discard the rest. In this study, we considered equal relative contribution of the cluster size and score in computation of the ranking criterion and we set the threshold on each individual score map that resulted in at most $n$=10 detected clusters. 

For the performance analysis, a cluster is considered a true detection if it overlaps at least 1 voxel of the ground truth annotation.

\bibliographystyle{cas-model2-names}

\bibliography{mybibliography}



\end{document}